# Equilibrium distribution and diffusion of mixed hydrogen-methane gas in gravity field

Shiyao Peng[1], Qiao He[2], Ducheng Peng[3*], Xin Ouyang[1], Xiaorui Zhang[4], Chong Chai[1], Lianlai Zhang[2], Xu Sun[5], Huiqiu Deng[6], Wangyu Hu[3], Jie Hou[3*]

1 PipeChina Institute of Science and Technology, Langfang 065006, China

2 CPECC Beijing Design Company, Beijing, 100085, China

3 College of Materials Science and Engineering, Hunan University, Changsha 410082, China

4 Beijing Gas and Heating Engineering Design Institute Co., Ltd., Beijing, 100032, China

5 National Engineering Laboratory for Pipeline Safety/MOE Key Laboratory of Petroleum Engineering, China University of Petroleum-Beijing, 102249, China

6 School of Physics and Electronics, Hunan University, Changsha 410082, China

**Abstract**

Repurposing existing natural gas pipelines is a promising solution for large-scale transportation of mixed hydrogen-methane gas. However, it remains debatable whether gravitational stratification can notably affect hydrogen partial pressure in the gas mixture. To address this issue, we combined molecular dynamics simulation with thermodynamic and diffusion theories. Our study systematically examined the equilibrium distribution of hydrogen-methane mixtures in gravity fields. We demonstrated that partial pressures of both gases decrease with altitude, with hydrogen showing slower decrease due to its smaller molar mass. As a result, the volume fraction of hydrogen is maximized at the top end of pipes. The stratification is more favorable at low temperature and large altitude drops, with notable gas stratification only occurring at extremely large drops in altitude, being generally negligible even at a drop of 1500 m. Furthermore, we showed that the diffusion time required to achieve the equilibrium distribution is proportional to gas pressure and the square of pipeline height. This requires approximately 300 years for a 1500 m pipeline at 1 bar. Therefore, temporary interruptions in pipeline gas transportation will not cause visible stratification. Our work clarifies the effect of gravity on hydrogen-methane gas mixtures and provides quantitative insights into assessing the stratification of gas mixtures in pipelines.

## 1. Introduction

The European energy crisis has refocused global attention on the issue of energy. Continued economic growth has led to a new peak in carbon emissions [1], making the transition to cleaner energy has become a matter of great urgency. As part of the European Green Deal, the EU has proposed that hydrogen as the optimal choice for achieving a carbon-neutral economy by 2050 [2]. Hydrogen has advantages such as being carbon-free and renewable [3], and can easily achieve large-scale production and application in comparison to solar, wind, or tidal power [4], making hydrogen energy a desirable clean energy source. Apart from being a new type of clean energy, hydrogen energy is also a form of conversion for other energy sources [5], which can alleviate long-distance transmission loads and large-scale power storage difficulties for wind energy and solar energy [6]. With the continuous development of hydrogen production technology, many countries and organizations (US [7, 8], EU[9] and China[10], etc.) have begun to carry out large-scale production and transportation of hydrogen .

---

[*] Corresponding author. E-mail address: pdc1277054663@hnu.edu.cn; jiehou@hnu.edu.cn;

Despite these advantages of hydrogen energy, building new hydrogen pipelines for large-scale transportation can be expensive and time-consuming. A promising solution is to using existing natural gas pipelines for transporting blended hydrogen-methane gas [11]. This approach is more cost-effective than building new hydrogen pipelines, as existing natural gas pipelines can be repurposed for transporting hydrogen with relatively minor modifications. However, this involves long-term exposure of pipeline steels to pressurized hydrogen gases, which can alter the mechanical properties of the pipelines and cause crack propagation, often leads to the reduction in fracture toughness, namely, "hydrogen embrittlement (HE) [12]. Numerous results from the adaptability evaluation of pipeline materials and connections indicate that hydrogen has a detrimental effect on them[13]. With an increase in the hydrogen mixing ratio, the hydrogen embrittlement sensitivity of pipeline materials and connections also increases, and the deterioration of the performance of pipeline connections becomes more pronounced. Zhang et al [14]. pointed out that the degree of damage caused by hydrogen to high strength pipeline steel is sensitive to the partial pressure. The fracture toughness and fatigue life of the high strength steel decrease with the increasing of hydrogen pressure [15-17]. Note methane is much denser than hydrogen gas, which raise the concern that whether gravity will cause notable fluctuations in the partial pressure of hydrogen in hydrogen-methane mixture.

Though many previous works have studied gas mixtures in gravity, the stratification behavior of hydrogen-methane mixture remains much debated. Azatyan et al. used the Boltzmann distribution to demonstrate that [18] the stratification of the mixed hydrogen-propane due to gravity can be ignored, even in the absence of convection. Badino et al. [19] point out that stratification may exist in non-mixing atmospheres with a drop of several kilometers, while at scales of several meters, almost no changes can be detected. Pitts et al. [20] conducted experiments on releasing hydrogen in a garage, and reported that measurements taken along a vertical array indicated a nearly uniform hydrogen volume fraction from top to bottom. These studies showed that gas generally does not stratify in gravity field. Nonetheless, a systematical investigation of hydrogen-methane mixture stratification behavior is yet to be made, and quantitative evaluations of pipeline height, temperature, mixing ratio, and diffusion time on the gas distribution remains missing.

Apart from these researches that claimed no stratification, there are also studies reported contradictory results with evident gas stratification. Liu et al. [21] measured the degree of hydrogen embrittlement to characterize hydrogen concentration in hydrogen-methane mixture, and modelled the evolution of the mixture in gravity with computational fluid dynamics (CFD) simulation, both approaches led to the conclusion of evident stratification. Nonetheless, it remains questionable that whether hydrogen concentration can be accurately backtracked from the degree of hydrogen embrittlement, and the CFD model used does not include Brownian movements of gas molecules which is the main contributor that blends the mixture. Shebeko et al. [22] conducted hydrogen leakage and diffusion experiments inside a vessel filled with quiescent air, showed that after stopping the injection for 250 min, the volume fraction of the hydrogen at the top and the bottom differed by 10%. However, as the diffusion of gases is a relatively slow process [23, 24], the it's possible that the stratification is simply a result of the initial non-uniform mixing state. In general, the stratification of hydrogen-methane mixture remains to be a controversial topic, which necessitates further systematical and quantitative investigations on the stratification behavior.

In this work, we addressed this issue by combining molecular dynamic simulations, Boltzmann distribution theory, and Mason-Weaver equations to analyze the equilibrium distribution and diffusion of hydrogen-methane mixtures in a gravity field. We systematically investigated the effects of height, temperature, and mixing ratio on the gas distribution. Our results demonstrate that significant gas stratification occurs only at extremely large drops in altitude, and the stratification at a drop of 1500 m is negligible. Additionally, we discovered that gas diffusion is generally slow, and it may take up to 300 years to achieve equilibrium distribution in a 1500 m pipeline at 1 bar. Thus, temporary interruptions in gas transportation through the pipeline will not cause visible stratification. Our study provides quantitative insights into the impact of gravity on hydrogen-methane gas mixtures, offering useful references for material selection and safety assessment of pipelines.

## 2. Method

In this work, we combined numerical simulations with analytical theory to investigate the behavior of hydrogen-methane mixture. We first carried out molecular dynamic (MD) simulations to direct model the movement of gas atoms/molecules in gravity field, then compare these simulations with statistic mechanics theory based on Boltzmann distribution. We also calculated the time evolution of gas distribution based on diffusion theories. Details of the numerical/analytical methods used are given in below.

### 2.1 Molecular dynamic simulations

All MD simulations in this work were carried out using the Large-scale Atomic/Molecular Massively Parallel Simulator (LAMMPS) package [25]with the results visualized using the OVITO software [26]. Note according to previous studies of equation of states of hydrogen and methane [27, 28], the behavior of $H_2$ and $CH_4$ is almost identical to ideal gas when in P<10 MPa region, indicating that the inter-atomic/molecular interaction is not important in our simulations. Therefore, we simply used He molecules as a modelling element for simplicity, and manually adjust its molar mass to 2 and 16 to mimic $H_2$ and $CH_4$ molecules. The embedded atom method (EAM) potential developed by Chen et al. [29] was adopted to describe He-He interactions. Two orthogonal simulation cells with dimensions $x \times y \times z = 20 \times 20 \times 400 \ nm^3$ and $20 \times 20 \times 4000 \ nm^3$ were adopted to model the effect of different heights (referred to as "400 nm box" and "4000 nm box" in below).

At the beginning of the simulation, 2000 H$_2$ + 2000 CH$_4$ atoms were randomly inserted into the 400 nm box (20000 H$_2$ + 20000 CH$_4$ for the 4000 nm box), which roughly corresponds to the number density of ideal gas at 300 K and 1 bar. After the insertion, the system was relaxed for 10 ps with isobaric-isothermal ensemble (i.e., NPT with P=1 bar and T=300 K) and with periodic boundary conditions applied to all three directions. The Nosé–Hoover thermostat was adopted and a timestep of 1 fs was used for all simulations. After reaching equilibrium, we switched to canonical ensemble (i.e., NVT with T=300 K), then turned on a gravity field along z direction, and applied a reflective boundary condition at the bottom and top of z direction to mimic a pipe with finite height. Finally, the system was relaxed under the gravity field for 2 ns, which is long enough for to establish new equilibrium in the gravity field. The equilibrium concentration profile of H$_2$ and CH$_4$ is calculated by analyzing the last 10 snapshots of the MD trajectory with an interval of 20 ps.

## 2.2 Boltzmann distribution

Considering a rigid vertical pipe filled with ideal gases, the normalized probability density of finding a gas molecular of type $i$ ($i$ = H$_2$ or CH$_4$ in this case) at height $h$ is given by the Boltzmann distribution:

$$\rho_i(h) = \frac{\exp\left(-\frac{m_i g h}{kT}\right)}{\int_0^H \exp\left(-\frac{m_i g h}{kT}\right) dh}, \#(1)$$

where $m_i$ is the mass of the gas molecular, $g$ is gravitational acceleration, $k$ is Boltzmann constant, $T$ is the absolute temperature, and $H$ is the maximum height of the pipe.

With the probability density given, assuming idea gas equation of state, we can calculate the distribution of partial pressure of gas by:

$$P_i(h) = P^{tot} F_i^{tot} \rho_i(h) H, \#(2)$$

where $P^{tot}$ is the average pressure of the whole system, $F_i^{tot}$ is the volume fraction of gas $i$ in the whole system. Below we find it's more convenient to consider the normalized partial pressure:

$$p_i(h) = P_i(h)/P^{tot} = F_i^{tot} \rho_i(h) H, \#(3)$$

Beside the normalized partial pressure, it's often useful to calculate the distribution of volume fraction of gas $i$:

$$F_i(h) = \frac{F_i^{tot} \rho_i(h) H}{\sum_j F_j^{tot} \rho_j(h) H} = \frac{p_i(h)}{\sum_j p_j(h)}, \#(4)$$

## 2.3 Mason-Weaver equation

Note the standard Fick's law of diffusion does not include the effect of external force field. To take gravity field into account, we can write the chemical potential of gas $i$ as:

$$\mu_i(h) = m_i g h + kT \ln p_i(h) + \mu_{ref}, \#(5)$$

where $\mu_{ref}$ is the chemical potential of an arbitrary reference state, which will be canceled out in the following calculation. Based on Eq. 5, we can calculate diffusion flux by:

$$J_i(h) = -\frac{Dp_i(h)}{kT}\frac{\partial \mu_i(h)}{\partial h} = -D\left[\frac{m_i g}{kT}p_i(h) + \frac{\partial p_i(h)}{\partial h}\right], \#(6)$$

where $D$ is the diffusion coefficient. Finally, we have the normalized partial pressure evolution against time:

$$\frac{\partial p_i(h)}{\partial t} = -\frac{\partial J_i(h)}{\partial h} = D\left[\frac{m_i g}{kT}\frac{\partial p_i(h)}{\partial h} + \frac{\partial^2 p_i(h)}{\partial h^2}\right]. \#(7)$$

Eq. 7 is also known as the Mason-Weaver equation [30], and can be solved analytically using the tool provided in Ref. [31]. In addition, it's easy to see that at long-time limit where $\frac{\partial p_i(h)}{\partial t} = 0$, the Mason-Weaver equation converges to the Boltzmann distribution that given by Eq. (1), demonstrating the consistence of two theories.

## 3. Results and discussion

### 3.1 Equilibrium distribution of hydrogen-methane mixture in gravity field

We started with direct MD simulations of 1:1 mixed $H_2$-$CH_4$ gases in different gravity field, and modeled their stratification behavior at atomic level. Fig. 1a shows the simulation results without any gravity, where we can see a uniform distribution of both $H_2$ and $CH_4$ molecules with no stratification observed, which is well expected due to the lack of gravitational separation. Next, to examine the effect of gravity on molecular distributions at nanometer scales, we applied a very large gravity field of $g = 10^{11} G$ (where $G = 9.8\ m^2/s$), which leads to a visible stratification as demonstrated in Fig. 1b, with $CH_4$ molecules showing apparent enrichment at the bottom of the simulation box. Note the distribution of $H_2$ molecules remains relatively uniform at $10^{11}$ G, indicating that gravity has much weaker effect on $H_2$ distribution due to its small molar mass. As we continue to increase the gravity to $10^{12}$ G, the stratification phenomenon in the pipeline becomes more evident (see Fig. 1c), with almost all $CH_4$ molecules segregated at z<100 nm region and a visible enrichment of $H_2$ molecules at the bottom region.

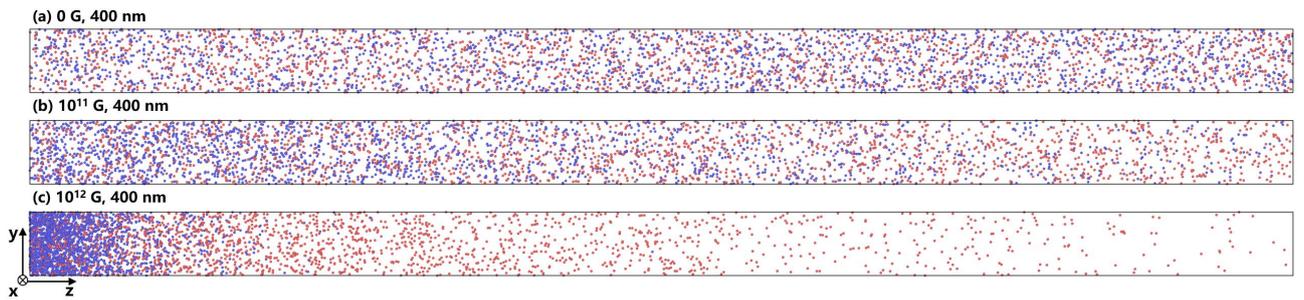

Figure 1. Snapshots of $H_2$-$CH_4$ mixtures at the end of MD simulations using the 400 nm box. (a-c) are results with 0, $10^{11}$, $10^{12}$ G gravity fields applied along z axis. Blue dots are $CH_4$ molecules and red dots represent $H_2$ molecules.

To obtain more quantitative insights against the above results, we calculated the normalized partial pressure and volume fraction profiles for $H_2$ and $CH_4$ molecules in the MD simulations in $10^{11}$ G gravity field. The results are plotted in Fig. 2a and 2f with a comparison to Boltzmann distribution predictions given by Eqs. 1-4, where our MD simulations matches perfectly with the Boltzmann distribution, demonstrating the validity of both approaches. We can see in Fig. 2a that in the $10^{11}$ G gravity field, $p_{CH_4}$ changes from ~137% (at z=0) to ~11% (at z=400 nm), while $p_{H_2}$ only changes from ~58% (at z=0) to ~43% (at z=400 nm). Both gases are enriched at the bottom and depleted at the top due to the gravitational effect. However, the volume fraction of $H_2$, $F_{H_2}$, shows a monotonic increase with the height (see Fig. 2d).

The increase of $H_2$ volume fraction with height is often interpreted as a risk for hydrogen embrittlement at the top region for a pipe containing mixed $H_2$-$CH_4$ gases. But here we should highlight that the increase of $F_{H_2}$ does not come from the increase of $p_{H_2}$, but from the more evident decrease of $p_{CH_4}$ (see Fig. 2a). While in terms of normalized partial pressure, both $p_{CH_4}$ and $p_{H_2}$ decreases with height. Since the hydrogen embrittlement risk is actually associated with the partial pressure of $H_2$, the embrittlement risk at the top region is actually lower than the bottom region due to gravity effects.

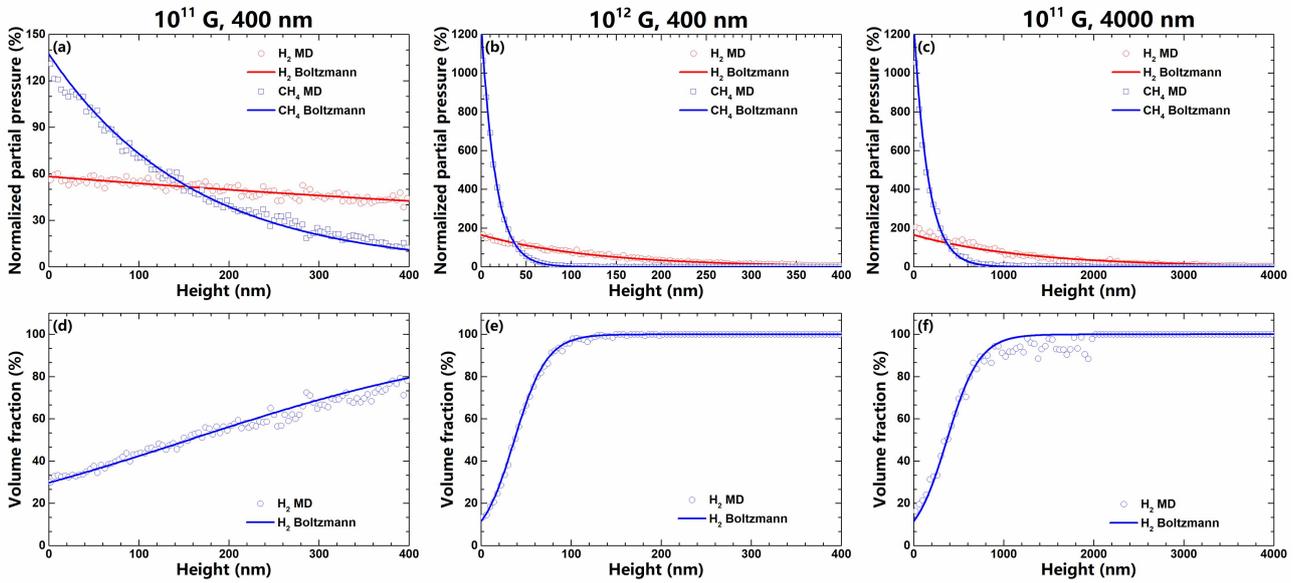

Figure 2. Normalized partial pressures of $H_2$ and $CH_4$ in (a) $10^{11}$ G gravity field with the 400 nm box, (b) $10^{12}$ G gravity field with the 400 nm box, and (c) $10^{11}$ G gravity field with the 4000 nm box. (d-f) are corresponding volume fraction. Hollow symbols are MD simulation results, lines are Boltzmann distribution predictions.

Next, we did similar calculations for gas distribution in $10^{12}$ G gravity field. As demonstrated in Fig. 2b and 2e, both $H_2$ and $CH_4$ gases show evident non-uniform distributions in the $10^{12}$ G gravity field, with MD simulation results also agree well with Boltzmann distribution. Here we find that $p_{CH_4}$ changes from ~1264% (at z=0) to ~0% (at z=400 nm), and $p_{H_2}$ changes from ~165% (at z=0) to ~7% (at z=400 nm). Comparing with that shown in $10^{11}$ G case, the segregation of gases is more evident in the $10^{12}$ G gravity field.

Note according to Eq. 1, the gas distribution profile is determined by the product of gravitational acceleration and height, namely, $g \times h$. In MD simulations, we can only handle very small heights (nm-$\mu$m) due to computational limitations, which is why a very strong gravity field ($10^{11}$ -$10^{12}$ G) is necessary to observe any visible gravity effect. Based on Eq. 1, we expect the distribution profile remain unchanged if we increase $h$ and decrease $g$ by a same factor (i.e., keep $g \times h$ constant). To examine this equivalency, we performed additional MD simulations using a 4000 nm simulation box and a $10^{11}$ G gravity field. The results are shown in Fig. 2c and 2f, where we find the normalized partial pressure and volume fraction is practically identical to that in Fig. 2b and 2e (except with different height scales) for both gases. Suggesting that our results can reflect the distribution at larger $h$ and lower $g$. In this way, the results shown in Fig. 2a and 2b should be equivalent to the distribution in 40 and 400 km of drop at 1 G, which means it takes a 100 km level of drop in altitude to observe evident stratification of $H_2$ and $CH_4$.

### 3.2 Effect of height, temperature, and mixing ratio on the stratification

The above results demonstrated that the equilibrium gravitational stratification of $H_2$-$CH_4$ gas mixture can be accurately predicted by the Eqs. 1-4. Below we further investigate different factors that could affect the stratification behavior. To begin with, we consider the realistic gravity field of 1 G and a pipe height of 1500 m (approximately the maximum drop in Shan–Jing pipeline) filled with 1:1 mixed $H_2$-$CH_4$ gases. Fig. 3 shows the equilibrium distribution calculated using Eqs. 1-4. Here we find that within the drop of 1500 m, normalized partial pressure vary almost linearly with the height. Similar to the above, both $p_{CH_4}$ and $p_{H_2}$ decreases with the height, with $p_{CH_4}$ decreasing faster than $p_{H_2}$ (see Fig. 3a), which in turn increase volume fraction of $H_2$, $F_{H_2}$, in the top region (see Fig. 3b). In general, large altitude drops promote the gravitational stratification of gases. But even with a large drop of 1500 m, the $p_{H_2}$ only changes slightly (from 49.7% to 50.3%), which is apparently too small to cause any notable impact on hydrogen embrittlement behavior.

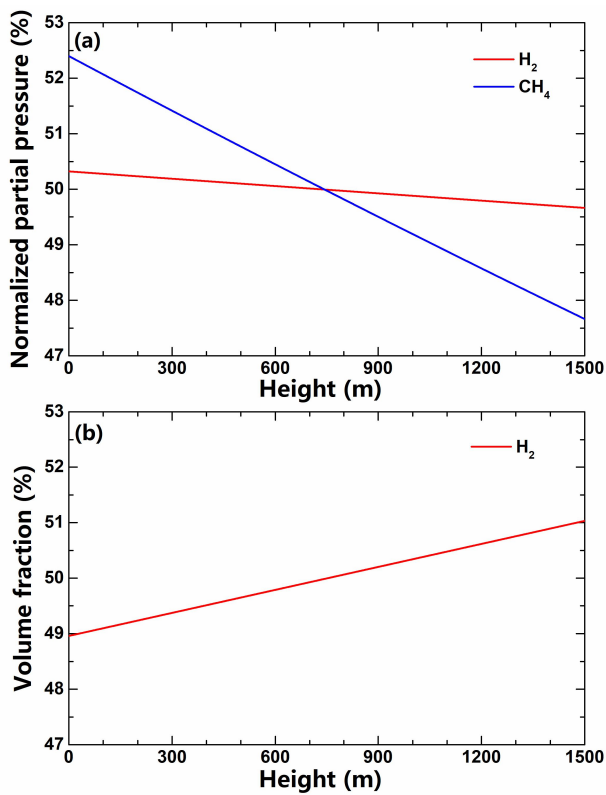

Figure 3. Equilibrium distribution profile of 1:1 mixed $H_2$-$CH_4$ gases within a 1500 m high pipe at 1 G and 300 K. (a) shows normalized partial pressures of two gases and (b) shows the volume fraction of $H_2$.

Next we examine the effect of temperature on the stratification behavior. Considering that partial pressures vary almost linearly against the height at km ranges, here we only focus on the partial pressure at the top and bottom of the pipe. Fig. 4 shows the calculated normalized partial pressure and volume fraction at different temperatures, where we find the difference between top and bottom partial pressures diminish with increasing temperature, suggesting that gas stratification is suppressed at elevated temperatures due to intensified Brownian motion of molecules. Yet even at a very low temperature of 150 K where gravitational stratification is more favorable, the difference between top and bottom $p_{H_2}$ is still insignificant (~1.2%).

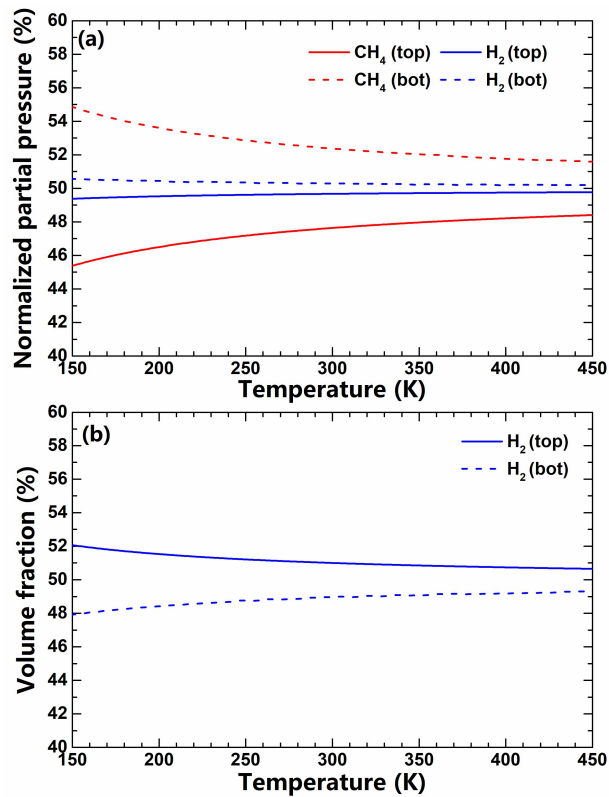

Figure 4. Normalized partial pressure and volume fraction of $H_2$ and $CH_4$ at the top and bottom of a 1500 m high pipe at 1 G, plotted as a function of temperature, where $H_2$ and $CH_4$ are 1:1 mixed. (a) shows normalized partial pressures of two gases and (b) shows the volume fraction of $H_2$.

Fig. 5 demonstrates the effect of mixing ratios (noted by overall $H_2$ volume fraction) on gas stratification behavior. We can see in Fig. 5a that normalized partial pressures of two gases are linearly correlated with their respective average concentration. That is, the difference between top and bottom $p_{H_2}$ is maximized in pure $H_2$ conditions. While the difference between top and bottom $F_{H_2}$ behaves differently (see Fig. 5b), as it first increases with average $p_{H_2}$, peaks at 50%, then start to reduce with average $F_{H_2}$. For the case of 1500 m high pipe at 300 K, the maximum difference in $F_{H_2}$ is ~2.7%.

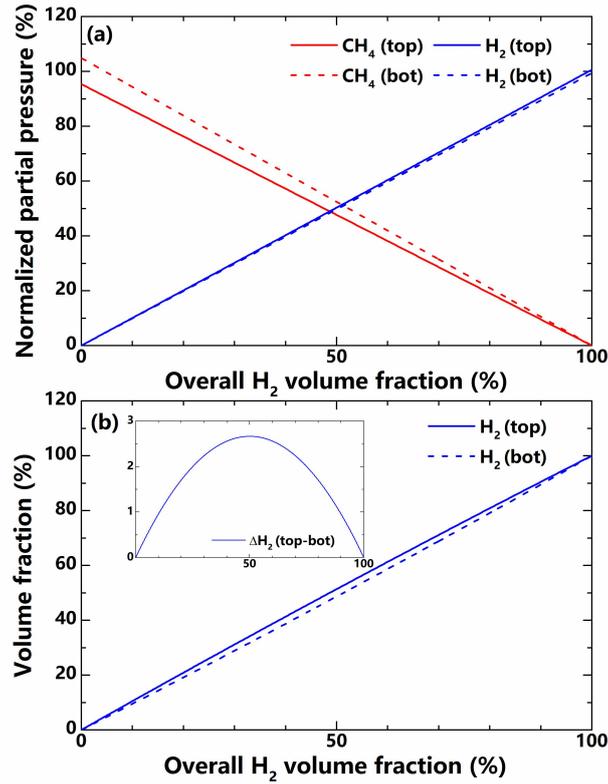

Figure 5. Normalized partial pressure and volume fraction of $H_2$ and $CH_4$ at the top and bottom of a 1500 m high pipe at 1 G and 300 K, plotted as a function of the overall $H_2$ volume fraction ($H_2$ mixing ratio). (a) shows normalized partial pressures of two gases and (b) shows the volume fraction of $H_2$.

### 3.3 Diffusion of hydrogen-methane mixture in gravity field

The above results reveal the behavior of mixed $H_2$-$CH_4$ gases at thermodynamic equilibrium. However, it remains unclear how long it takes to reach such equilibrium. To understand the diffusion behavior of the gas mixture, we fist calculated its diffusion coefficient according to [23, 24]:

$$D(P,T) = 3.13 \times 10^{-5} T^{1.765} \frac{0.1 \text{ MPa}}{P} \text{cm}^2\text{s}^{-1}. \#(8)$$

At $P = 0.1$ Mpa and $T = 300$ K, we have $D = 0.74$ cm$^2$s$^{-1}$. Now consider a pipe with height $H = 1500$ m, a rough estimation for the characteristic time would be t $\approx H^2/D \approx 242$ years. Which means the gas diffusion is a very slow process at the scale of km. To justify this estimation, we calculated the time evolution of $F_{H_2}$ profile by solving the Mason-Weaver equation (Eq. 7). The results are shown in Fig. 6, where we can see it takes about 300 years for fully blended gas mixtures to approach its $t \to \infty$ limit in the 1500 m high pipe (i.e., the equilibrium state in Fig. 3b). This time scale generally agrees with the characteristic time estimated above. Moreover, we noticed that within 1 month, the volume fraction profile remains virtually unchanged. That means even if the stratification is somehow thermodynamically feasible, it's still unlikely to observe any visible stratification during a temporary interruption in pipeline gas transportation.

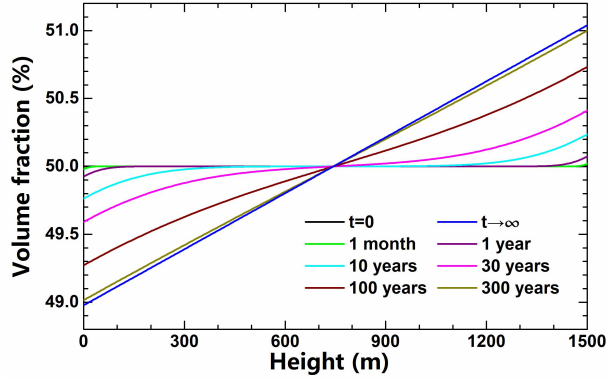

Figure 6. Evolution of volume fraction profile of $H_2$ in a 1500 m high pipe at 1 G and 300 K, where $H_2$ and $CH_4$ are 1:1 mixed and fully blended at t=0.

Fig. 7 presents more details on the diffusion behavior by visualizing $\Delta F_{H_2}$ at the top of pipes against time. In general, $\Delta F_{H_2}$ first increases linearly with $t^{1/2}$ (note the logarithm scale), then gradually plateaus after reaching thermodynamic equilibrium. As demonstrated in Fig. 7a, increasing the pressure leads to a shift in the curve along the time axis, which means the time required to reach steady state increases linearly with increasing pressure, being approximately $10^5$, $10^6$, and $10^7$ days under 0.1, 1, and 10 MPa. According to Eq. 8, the diffusion coefficient is inversely proportional to gas pressure. As a result, diffusion time will be proportional to the pressure, which explains the results shown in Fig. 7a. We also evaluated the effect of pipe height on the diffusion behavior with relative data plotted in Fig. 7b. Here we find that increasing height does not shift the curve, but postpones the time needed to establish equilibrium. Note this time is proportional to the square of the height, namely, the time required for reaching equilibrium in 1500 m pipe is 100 times higher than that in 150 m pipe. That means thermodynamic and kinetic feasibility of stratification usually cannot be satisfied simultaneously, as evident stratification requires a large pipe height, which in turn prevents the establishment of equilibrium in limited time.

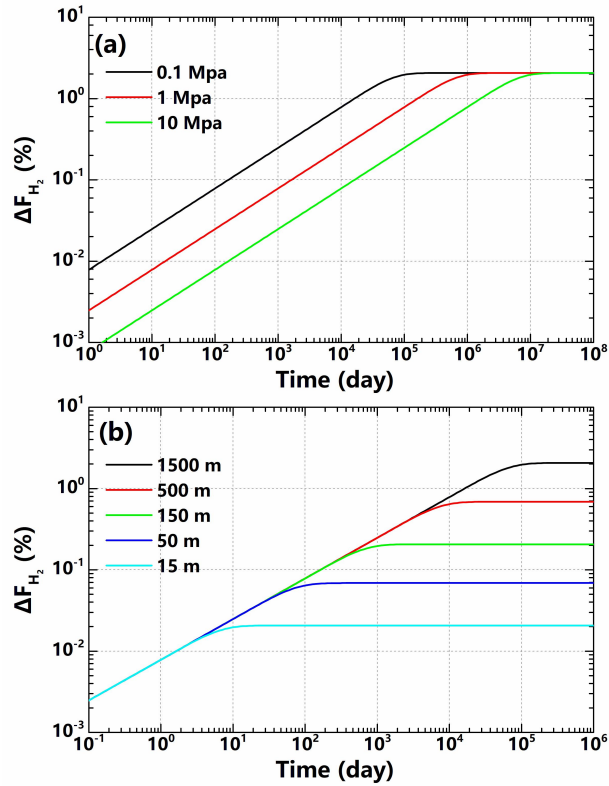

Figure 7. Change of volume fraction of $H_2$ at the top of pipes at 1 G and 300 K, where $H_2$ and $CH_4$ are 1:1 mixed and fully blended at t=0. (a) 1500 m high pipe with different pressures, (b) different pipe heights with 0.1 MPa pressure.

Note the above calculations are based on initially fully blended gas mixtures, and the diffusion of gases will lead to gravitational stratification. Here we consider another situation in which gas mixtures are initially fully stratified, with pure $H_2$ fills the top half and pure $CH_4$ fills the bottom half of a short pipe with 1.5 m of height. In this case, the equilibrium stratification is negligible inside the 1.5 m pipe, which means the initial stratification will diminish at $t \to \infty$. However, from the calculation results shown in Fig. 8, we find the stratification remain quite evident at 1 hour, and is still visible even after 10 days of gas diffusion. These results suggest that if gases were not well blended at the beginning, then it could lead to a misleading conclusion that gas can stratify due to gravity effect, which should be carefully checked in related experiments.

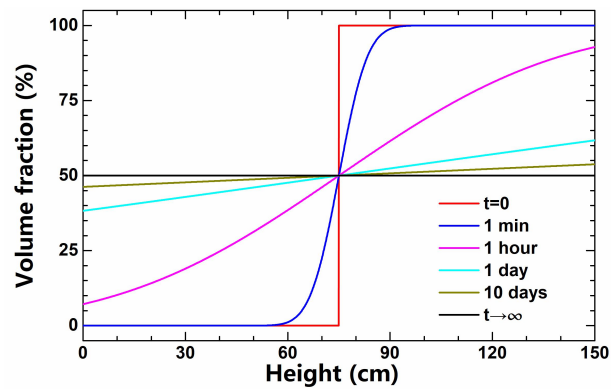

Figure 8. Evolution of volume fraction profile of $H_2$ in a 1.5 m high pipe at 1 G and 300 K, where $H_2$ and $CH_4$ are fully stratified at t=0, with pure $H_2$ fills the top half and pure $CH_4$ fills the bottom half.

## 4. Conclusion

In this study, we employed molecular dynamic simulations and analytical theories to investigate the equilibrium and diffusion behavior of mixed hydrogen-methane gas in a gravity field. Our findings led to the following conclusions:

1. The molecular dynamic simulations of $H_2$ and $CH_4$ gases in a gravity field show that gravitational stratification of these gases requires either an extremely strong gravity field or very large drops in altitude (on the order of 100 km), in agreement with Boltzmann distribution theory predictions.
2. Both $CH_4$ and $H_2$ lose their partial pressure at high altitudes. This effect is more evident in $CH_4$ due to its larger molar mass, which, in turn, maximizes the volume fraction of $H_2$ at the top end of pipes.
3. Large pipe height and low temperature promote the stratification. However, even with a low temperature of 150 K and a large height of 1500 m, the stratification is insignificant and will not significantly affect the risk of hydrogen embrittlement.
4. The diffusion time required to reach thermodynamic equilibrium increases linearly with pressure and the square of pipe height. It takes approximately 300 years to reach equilibrium in a 1500 m high pipe, and temporary interruptions in pipeline gas transportation will not cause significant stratification. However, if gases are not well blended initially, it can lead to the misleading conclusion that gas can stratify in a gravity field.

This present study provides quantitative insights into evaluating the stratification of gas mixtures in pipelines and sheds light on the equilibrium and diffusion behavior of mixed hydrogen-methane gas in a gravity field.

**Data availability**

The data generated and/or analysed within the current study will be made available upon reasonable request to the authors.

**Acknowledgement**

This work was financially supported by Research on key technologies of hydrogen-mixed natural gas transportation by pipeline in service (PipeChina, project No: DTXNY202203), Pilot scale study on green hydrogen production by alkaline water electrolysis of hundred Nm³/h capacity (CNPC, project No.: 2022DJ5006(GF)), and by Research and development of key technologies for medium